\documentclass[12pt,twoside]{article}
\usepackage{fleqn,espcrc1}
\usepackage{tabularx}
\usepackage{graphicx,epsfig}
\usepackage[figuresright]{rotating}

\newcolumntype{Y}{>{\small\raggedleft\arraybackslash}X}

\newcommand{\AmS}{{\protect\the\textfont2
  A\kern-.1667em\lower.5ex\hbox{M}\kern-.125emS}}
\newcommand{\st}{\scriptstyle}

\newcommand{\cl}{\centering}

\title{{\small UNITU-THEP-7/1999 \hfill hep-ph/9907563}\\~\\
Octet and Decuplet Baryons \\
in a Confining and Covariant Diquark-Quark Model\footnote{Supported by the 
BMBF (06--TU--888)
and by the DFG (We 1254/4-1).\\
Talk given by R.~Alkofer at PANIC 99.} }

\author{R.~Alkofer,
 S.~Ahlig, 
C.~Fischer, 
M.~Oettel, 
and H.~Reinhardt}

\begin{document}

\maketitle \noindent
Institut f\"ur Theoretische Physik der Universit\"at T\"ubingen\\
Auf der Morgenstelle, 72076 T\"ubingen, Germany\\

\begin{abstract}
We treat baryons as bound states of scalar or axialvector diquarks and
a constituent quark which interact through quark exchange. We obtain fully four-dimensional
wave functions for both octet and decuplet baryons as solutions of the corresponding Bethe-Salpeter
equation. Applications currently under investigation are: electromagnetic and strong form factors and
strangeness production processes.
\end{abstract}

\section{Motivation}
Different types of hadronic models describe
various aspects of baryon physics. Among them are
nonrelativistic quark models, various sorts of bag models and approaches
describing baryons by means of collective variables like topological or non-topological
solitons \cite{Bhad88}. Most of these models
are designed to work in the low energy region and generally do not match the calculations within perturbative 
QCD. Considering the great
experimental progress in the medium energy range \cite{Proc99}, 
there is a high demand for models describing 
baryon physics in this region that connects the low and high energy regimes. To make progress in this direction we investigate a covariant
formulation of a diquark-quark-model of baryons. 

Our motivation to choose such an approach is fed from two sources.
On the one hand, when starting with the fully relativistic
Faddeev equation for bound states of three quarks, 
diquarks appear as effective degrees of freedom.
These diquarks stand for correlated quark-quark pairs
inside baryons. Thus they should not be confused with the notion of diquark condensates in
the context of colour superconductivity. On the other hand,
diquarks as constituents of baryons are naturally obtained when one starts with
an NJL-type of  model of colour octet flavour singlet quark currents \cite{alk95}. 
Although in the limit $N_c \rightarrow \infty$ baryons emerge as solitons of meson 
fields \cite{alk96},
it can be shown for the case of three colours that both effects, binding through quark exchange in
the diquark-quark picture and through mesonic effects, contribute equally \cite{zuc96}.

\section{Solving the four-dimensional Bethe-Salpeter equation: masses and wave functions}

Starting from the Faddeev equation, one can approximate the two-quark 
irrreducible $T$-matrix by separable contributions that can be viewed as
loosely bound diquarks. The full three-body problem reduces then to a two-body 
one, in which bound states appear as the solution of a homogeneous 
Bethe-Salpeter equation. The attractive interaction between quark and 
diquark is hereby provided by quark exchange.
In \cite{oett98} we formalize this procedure by an effective Lagrangian
containing constituent quark, scalar diquark and axialvector diquark fields.
This leads to a coupled set of Bethe-Salpeter equations for octet and
decuplet baryons.

\begin{table}[t]
\caption{Components of the octet baryon wave function with their
respective spin and orbital angular momentum. 
$(\gamma_5 C)$ corresponds to
scalar and $(\gamma^\mu C),\,\mu=1 \dots 4,$ to axialvector
diquark correlations. Note that the partial waves in the first row
possess a non-relativistic limit. See \cite{oett98} for
further details.}
\label{wave}
\begin{tabular}{p{0.1cm}p{2.7cm}cccc}\hline
 &&&&& \\
\multicolumn{2}{l}{\parbox{2.8cm}{ { \mbox{``non-relativistic''} partial waves} }  } &
  $\pmatrix{ \st \chi \cr \st 0 } \st{(\gamma_5 C)}$    &
  $\st\hat P^4{\pmatrix{ \st 0\cr \st \chi}} (\gamma^4 C)$    &
  ${\pmatrix{\st i\sigma^i\chi \cr \st 0}} \st (\gamma^i C)$    &
  ${\pmatrix{\st i\left(\hat p^i(\vec{\sigma}\hat{\vec{p}})-\frac{\sigma^i}{3}\right) \chi\cr \st 0}} \st (\gamma^i C)$ \\  
 &  {spin} & {1/2} & {1/2} & {1/2} & {3/2} \\
 &  {orbital angular momentum} & { $s$} & {$s$} & {$s$} & {$d$} \\
\multicolumn{2}{l}{\parbox{2.8cm}{ { ``relativistic'' \mbox{partial waves} }  }} &
  $\pmatrix{ \st 0 \cr \st \vec \sigma \vec p \chi } \st (\gamma_5 C)$    &
  $\st \hat P^4{\pmatrix{ \st (\vec{\sigma}\vec{p})\chi\cr \st 0}}(\gamma^4 C)$    &
  ${\pmatrix{\st 0\cr \st i\sigma^i(\vec{\sigma}\vec{p})\chi}}\st (\gamma^i C)$    &
  ${\pmatrix{\st 0\cr \st i\left(p^i-\frac{\sigma^i(\vec{\sigma}\vec{p})}{3}\right)\chi}} \st (\gamma^i C)$ \\   
 &  {spin} & \cl{1/2} & \cl{1/2} & \cl{1/2} & {3/2} \\
 & {orbital angular momentum} & \cl{ $p$} & \cl{$p$} & \cl{$p$} & {$p$}\\ \hline
\end{tabular}
\end{table}

We avoid unphysical thresholds by an effective parameterization of
confinement in the quark and diquark propagators. 
We then solve the complete four-dimensional Bethe-Salpeter equation
in ladder approximation and obtain wave functions for the octet and
decuplet baryons \cite{oett98}.  
The Lorentz invariance
of our model has been checked explicitly by choosing different frames.

The implementation of the appropriate Dirac  and Lorentz
representations 
of the quark and diquark parts of the wave functions leads to a unique
decomposition in the rest frame of the baryon. Besides the well known $s$-wave and
$d$-wave components of non-relativistic formulations of the baryon octet
we additionally obtain
non-negligible $p$-wave contributions which demonstrates again the need for
covariantly constructed models. Table \ref{wave} summarizes the
structure of the octet wave function. Each of the eight components is to be 
multiplied with a scalar function which is given in terms of an expansion
in hyperspherical harmonics and is computed numerically.

In order to obtain the mass spectra for the
octet and decuplet baryons we explicitly break SU(3) flavour symmetry by
a higher strange quark constituent mass.
Using the nucleon and the delta mass as input our calculated mass spectra \cite {oett98} are in good
agreement with the experimental ones, see Table \ref{masses}.
The wave functions for baryons with distinct
strangeness content but same spin differ mostly due to
flavour Clebsch-Gordan coefficients, the respective invariant functions being
very similar.
Due to its special
role among the other baryons, we investigated the
$\Lambda$ hyperon in more detail and discussed its vertex amplitudes.
In our approach, the $\Lambda$ acquires a small flavour singlet
admixture which is absent in $SU(6)$ symmetric
non-relativistic quark models.

\begin{table}
\caption{Octet and decuplet masses obtained with two different parameter sets.
Set I represents a calculation with weakly confining propagators, Set II
with strongly confining propagators, see \cite{oett98}.
All masses are given in GeV.}
\begin{tabularx}{\linewidth}{YYYYYYYYY} \hline
 & $m_u$ & $m_s$ & $M_\Lambda$ & $M_\Sigma$ & $M_\Xi$ & $M_{\Sigma^*}$ & $M_{\Xi^*}$ & $M_\Omega$ \\ \hline
 &&&&&&&& \\
Set I & 0.5 & 0.65 & 1.123 & 1.134 & 1.307 & 1.373 & 1.545 & 1.692 \\
Set II& 0.5 & 0.63 & 1.133 & 1.140 & 1.319 & 1.380 & 1.516 & 1.665 \\
Exp.  &     &      & 1.116 & 1.193 & 1.315 & 1.384 & 1.530 & 1.672 \\ 
 &&&&&&&& \\ \hline
\end{tabularx}
\label{masses}
\end{table}

\section{Applications: Form Factors and Strangeness Production}

A significant test and a first application of our model is the calculation of
various form factors \cite{hell97,ahlig99}. The most important ingredient are the
fully four-dimensonal wave functions described above. It turns out that already the electromagnetic
form factors of the nucleon provide severe resctrictions for the parameters of the model.

The pion-nucleon form factor is calculated in view of its possible use in a spectator model
for nucleon-nucleon scattering processes. At the soft point, $Q^2=0$, we find
good agreement with experiment. In the spacelike region our $g_{\pi NN}$ falls like a 
monopole with a large cutoff used also in One-Boson-Exchange (OBE) models.
Compared with a calculation including only scalar diquarks \cite{hell97} we find a
lower value for the pion-nucleon coupling at the soft point.
Serving as a central ingredient for strangeness production processes the kaon-nucleon-lambda
form factor $g_{KN\Lambda}$ is an issue of special interest. Due to flavour algebra the 
isospin configuration of the $\Lambda$ singles out the scalar diquark as the only diquark
contributing to nucleon-lambda transitions. With a pseudoscalar kaon not coupling to
the scalar diquark we find ourselves in a comfortable position to handle such
transitions. Results can be found in a forthcoming paper \cite{ahlig99}.

Going beyond the calculation of form factors we work on the application of our approach
to production processes like kaon photoproduction and the associated strangeness production
close to threshold.
Great experimental progress in the last few years gave access to kinematical  as well
as some spin observables due to self-analyzing $\Lambda$ decay. We try to reproduce these
observables in our model with the aim to get further insight into the mechanisms of
strangeness production inside the nucleon. 

\begin{figure}[t]
\begin{center}
\epsfig{file=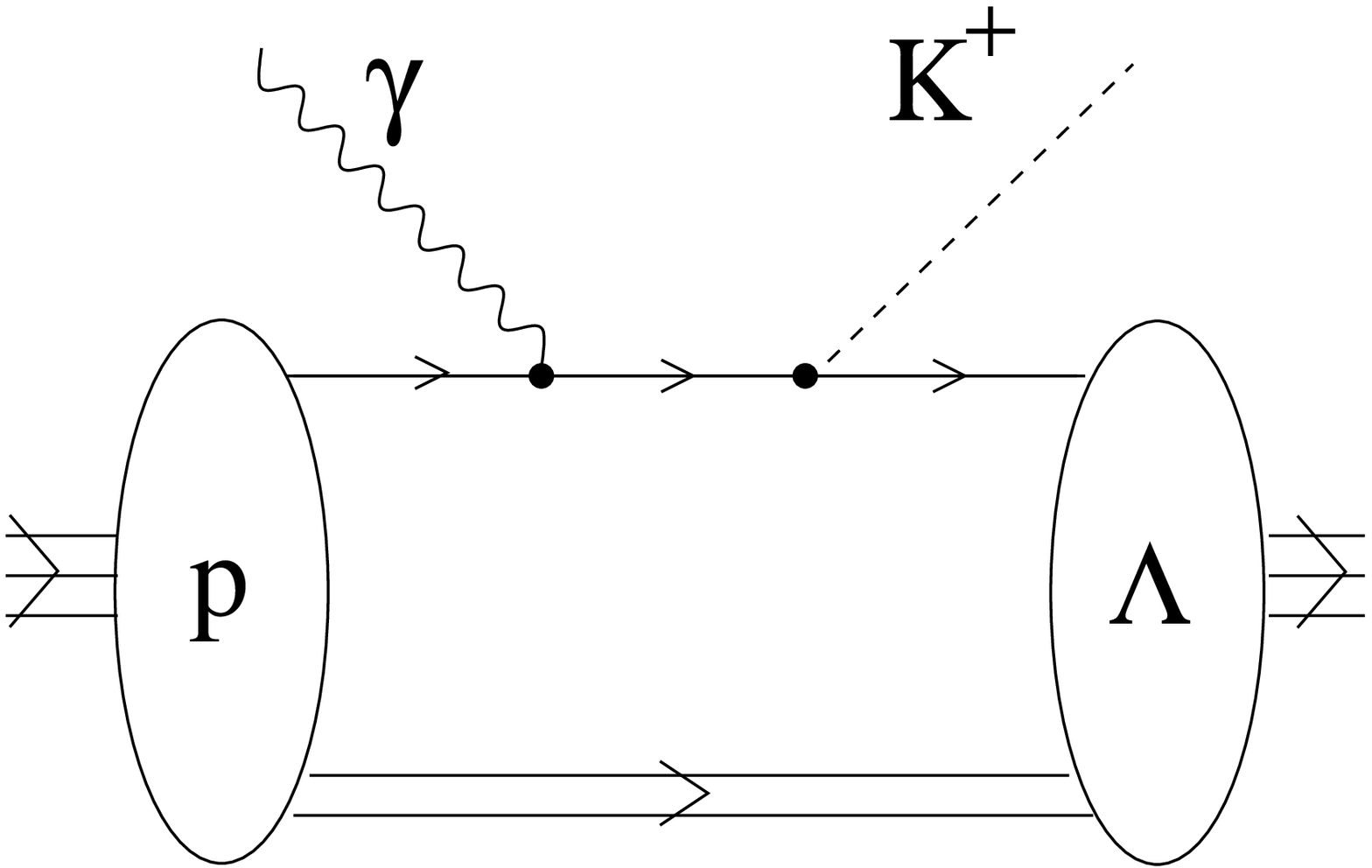,width=4.3cm} \hspace{1.6cm}
\epsfig{file=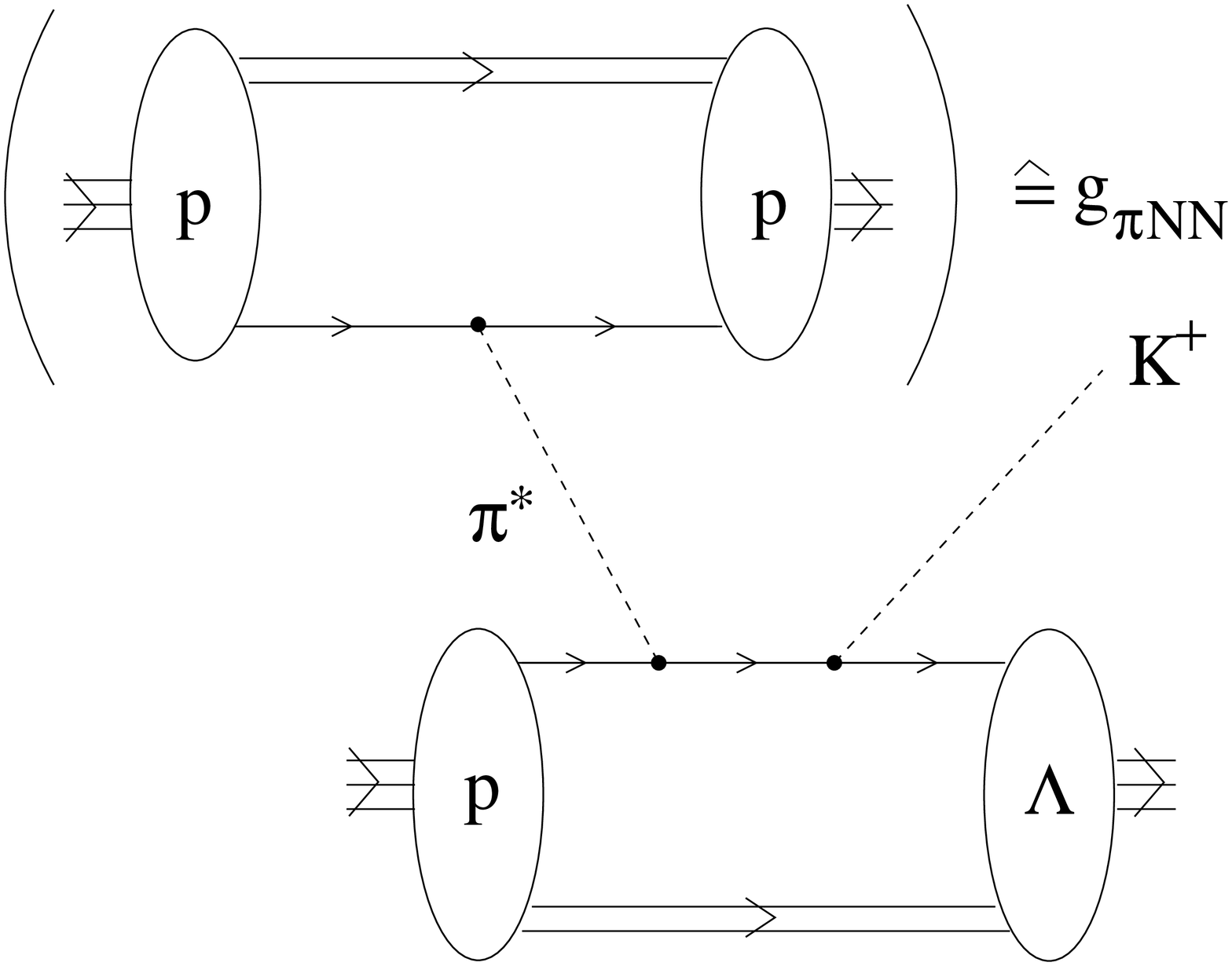,width=6.3cm}
\end{center}
\vspace{-5mm}
\caption{Typical diagrams to be computed in the diquark
spectator picture. Left: kaon photoproduction. Right:
strangeness production in proton-proton collisions.}
\label{pics}
\end{figure}

In the case of kaon photoproduction, $\gamma p \rightarrow K\Lambda$,
preliminary results are gained by computing the left diagram of fig.
\ref{pics} and the corresponding one with photon and kaon line crossed.
Although the outcome of the total cross section is encouraging, the 
result for the polarisation asymmetries of the $\Lambda$ falls too short. 

Additionally, we describe the process $pp \rightarrow pK\Lambda$ employing
an OBE-picture for internuclear forces. From our viewpoint, the diquarks act as mere
spectators, so that the exchanged bosons couple only to the valence quarks of the baryons.
In a first step we investigate the contributions stemming
from pseudoscalar kaon and pion interchange bet\-ween these two protons.
The latter process is depicted in the right diagram of fig. \ref{pics}.
If instead of the pion a kaon is exchanged, the upper part of this diagram
changes to the form factor $g_{KN\Lambda}$.

Possible applications are not exhausted by these examples. The diquark spectator picture can
be extended to, {\it e.g.}, real and virtual Compton scattering. The
quark structure functions (see also \cite{kusa97}), which are obtained
from the imaginary part of the virtual Compton scattering amplitude,
would allow a comparison with the results of perturbative calculations.
\\[0.2cm]
{\bf Acknowledgement:}

R. A. thanks  W. Bentz and P. Maris  for interesting discussions
and the organizers and conveners of PANIC 99.

\end{document}